\documentclass{aa}
\usepackage{times,float}
\usepackage{epsfig}
\usepackage{graphics}
\usepackage{amssymb}
\usepackage{curves}
\usepackage{eepic}
\usepackage{epic}

\begin{document}

\title{The short-duration GRB~050724 host galaxy in the context of the
  long-duration GRB  hosts\thanks{Based on observations  made with the
  Telescopio  Nazionale Galileo  (prop.  ID  TAC-1238) and  the Nordic
  Optical  Telescope, operated  on  the  island of  La  Palma, in  the
  Spanish Observatorio del Roque de  los Muchachos of the Instituto de
  Astrof\'{\i}sica de Canarias, and on observations carried out at the
  Centro Astron\'omico Hispano Alem\'an  (CAHA) at Calar Alto operated
  jointly  by  the  Max-Planck   Institut  f\"ur  Astronomie  and  the
  Instituto de Astrof\'{\i}sica de Andaluc\'{\i}a (CSIC).}}

\titlerunning{The host galaxy of the short GRB~050724}

\author{
        J.~Gorosabel\inst{1}
   \and A.J.~Castro-Tirado\inst{1}
   \and S.~Guziy\inst{1}
   \and A. de Ugarte Postigo\inst{1}
   \and D. Reverte\inst{1}
   \and A. Antonelli\inst{2}
   \and S. Covino\inst{3}
   \and D.~Malesani\inst{4}
   \and D. Mart\'{\i}n-Gord\'on\inst{1}
   \and A. Melandri\inst{2}
   \and M. Jel\'\i nek\inst{1}
   \and N. Elias de la Rosa\inst{5}
   \and O.~Bogdanov\inst{6}
   \and J.M.~Castro~Cer\'on\inst{7}
}
\institute{Instituto de Astrof\'{\i}sica de Andaluc\'{\i}a (IAA-CSIC),
  Apartado de Correos, 3.004, E-18.080 Granada, Spain.
  \and
  INAF-Osservatorio Astronomico di Roma, via di Frascati 33, I-00040 Monteporzio Catone (Rome), Italy.
  \and 
  INAF-Osservatorio Astronomico di Brera, via E. Bianchi 46, I-23807 Merate (LC), Italy.
  \and
  International School  for Advanced Studies  (SISSA-ISAS), via Beirut
  2-4, I-34014 Trieste, Italy.
  \and
  INAF-Osservatorio Astronomico di Padova, via dell'Osservatorio 8, I-36012 Asiago (VI), Italy.
  \and
  Astronomical Observatory of Nikolaev State University, Nikolskaja, 24, Nikolaev, 54030, Ukraine
  \and
Dark Cosmology Centre, Niels Bohr Institute, University of Copenhagen, Juliane Maries Vej 30, DK-2100 Copenhagen \O, Denmark.
}

\offprints{ \hbox{J. Gorosabel, e-mail:{\tt jgu@iaa.es}}}

\date{Received  / Accepted }

%%%%%%%%%%%%%%%%%%%%%%%%%%%%%%%%%%%%%%%%%%%%%%%%%%%%%%%%%%%%%%%%%%%%%%%%%%%%%%%

\abstract{We report optical  and near-infrared broad band observations
  of the short-duration GRB~050724  host galaxy, used to construct its
  spectral   energy   distribution  (SED).    Unlike   the  hosts   of
  long-duration  gamma-ray bursts (GRBs),  which show  younger stellar
  populations,  the SED  of the  GRB~050724 host  galaxy  is optimally
  fitted  with a  synthetic  elliptical galaxy  template  based on  an
  evolved stellar population (age $\sim2.6$ Gyr).  The SED of the host
  is difficult  to reproduce with  non-evolving metallicity templates.
  In contrast, if the short  GRB host galaxy metallicity enrichment is
  considered,   the   synthetic  templates   fit   the  observed   SED
  satisfactorily.   The internal  host extinction  is low  ($A_{\rm v}
  \lesssim  0.4$  mag) so  it  cannot  explain  the faintness  of  the
  afterglow.    This   short  GRB   host   galaxy   is  more   massive
  ($\sim5\times10^{10} M_{\odot}$)  and luminous ($\sim1.1 L^{\star}$)
  than most of the  long-duration GRB hosts.  A statistical comparison
  based  on the  ages of  short- and  long-duration GRB  host galaxies
  strongly suggests that short-duration GRB hosts contain, on average,
  older progenitors.   These findings  support a different  origin for
  short-  and  long-duration   GRBs.}

\maketitle

\section{Introduction}

It  is widely  accepted nowadays  that long-duration  gamma-ray bursts
(GRBs)  are  related  to   core-collapse  supernovae  (Stanek  et  al.
\cite{Stan03};   Hjorth  et   al.   \cite{Hjor03});   the   origin  of
short-duration GRBs  is, however, still a mystery  (see Kouveliotou et
al.   \cite{Kouv93} for  the definition  of short  GRBs).  Up  to 2005
efforts to  detect short-duration GRBs at  longer wavelengths remained
unfruitful  (see  Hurley  et  al.   \cite{Hurl02},  Gorosabel  et  al.
\cite{Goro02}, and  references therein), but with  prompt and accurate
localizations recently  reported for  four short-duration GRBs  by the
{\it Swift}  and {\it \mbox{HETE-2}}  satellites (GRB~050509B, Gehrels
et  al.   \cite{Gehr05};  GRB~050709,  Butler et  al.   \cite{Butl05};
GRB~050724, Covino  et al.  \cite{Covi05a}; and  GRB~050813, Retter et
al.   \cite{Rett05}),  the  field  is  currently  undergoing  a  rapid
breakthrough.

Interestingly  enough,  for the  above  mentioned four  short-duration
GRBs, relatively  bright galaxies ($18  \lesssim R \lesssim  24$) have
been  found close,  or  within, the  corresponding  X-ray error  boxes
(Bloom et  al.  \cite{Bloo05a}; Covino et  al.  \cite{Covi05b}; Hjorth
et al.   \cite{Hjor05}; Antonelli  et al.  \cite{Anto05};  Gladders et
al.   \cite{Glad05};  see  also  Prochaska  et  al.   \cite{Proc05a}).
Recent statistical  studies (Tanvir et al.   \cite{Tanv05}; Gal-Yam et
al.    \cite{Galy05a})  suggesting  an   excess  of   bright  galaxies
coincident with short GRB error boxes support this correlation.

\begin{table*}
\begin{center}
\caption{Optical and NIR photometric observations of the host of GRB~050724. The
magnitudes are not corrected for Galactic extinction.}
\begin{tabular}{lcccccccc}
\hline
{\small Telescope}&{\small  Filter}&{\small Date UT}&{\small T$_{\rm exp}$}&{\small  Seeing}& Effective  &{\small Bandpass}&{\small Magnitude}&{\small Magnitude}\\
{\small (+Instrument)}&            &{\small (mid exposure)   }&{\small (s)          }&{\small ($^{\prime \prime}$)}&{\small wavelength (\AA)}&{\small width (\AA)}&{\small (Vega)}&{\small (AB)}\\
\hline

{\small 2.5NOT (+ALFOSC)}&{\small $U$}&{\small   6.903/08/05 }&{\small 900}&{\small 1.3       }&{\small   3660}&{\small 355}&{\small $>$22.20$^{\dagger}$}&{\small  $>$23.09$^{\dagger}$}\\
{\small 2.5NOT (+ALFOSC)}&{\small $B$}&{\small   6.884/08/05 }&{\small 600}&{\small 1.3       }&{\small   4384}&{\small 700}&{\small 22.47$\pm$0.12}&{\small  22.34$\pm$0.12}\\
{\small 2.5NOT (+ALFOSC)}&{\small $V$}&{\small   6.921/08/05 }&{\small 600}&{\small 1.2       }&{\small   5368}&{\small 527}&{\small 20.69$\pm$0.05}&{\small  20.69$\pm$0.05}\\
{\small 2.5NOT (+ALFOSC)}&{\small $R$}&{\small   6.929/08/05 }&{\small 600}&{\small 1.5       }&{\small   6627}&{\small 768}&{\small 19.58$\pm$0.03}&{\small  19.83$\pm$0.03}\\
{\small 3.6TNG (+NICS)}&{\small $J_s$}&{\small   4.872/08/05 }&{\small 20$\times$60}&{\small 1.1}&{\small 12541}&{\small 964}&{\small 16.88$\pm$0.04$^{\star}$ }&{\small  17.83$\pm$0.04$^{\star}$}\\
{\small 3.6TNG (+NICS)}&{\small $J_s$}&{\small  12.887/08/05 }&{\small 25$\times$60}&{\small 1.0}&{\small 12541}&{\small 964}&{\small                          }&{}\\
{\small 3.5CAHA (+OMEGA$_{2000}$)}&{\small $J$}&{\small 25.868/07/05 }&{\small  30$\times$90}&{\small  2.5}&{\small  12861}&{\small  1713}&{\small  16.87$\pm$0.13$^{\ddag}$ }&{\small  17.84$\pm$0.13$^{\ddag}$}\\
{\small 3.6TNG (+NICS)}&{\small $H$  }&{\small   4.911/08/05 }&{\small 21$\times$60}&{\small 1.0}&{\small 16289}&{\small 173}&{\small 15.86$\pm$0.05          }&{\small  17.24$\pm$0.05}\\
{\small 3.6TNG (+NICS)}&{\small $K_s$}&{\small   4.892/08/05 }&{\small 21$\times$60}&{\small 0.9}&{\small 21203}&{\small 204}&{\small 15.03$\pm$0.04$^{\star}$}&{\small  16.86$\pm$0.04$^{\star}$}\\
{\small 3.6TNG (+NICS)}&{\small $K_s$}&{\small  12.863/08/05 }&{\small 25$\times$60}&{\small 1.1}&{\small 21203}&{\small 204}&{\small                         }&{}\\
\hline
\multicolumn{4}{l}{\small $\ddag$ Only used for checking the afterglow contamination}\\
\multicolumn{4}{l}{\small $\dagger$ 3 $\sigma$ upper limit}\\
\multicolumn{4}{l}{\small $\star$ Combined images from two epochs}\\
\hline
\label{table1}
\end{tabular}
\end{center}
\end{table*}

The short GRB~050724  was detected by {\it Swift} on  24 July 2005, at
12:34:09 UT. It showed an X-ray afterglow consistent with the location
of a bright  galaxy present on the Digitized  Sky Survey (Antonelli et
al.  \cite{Anto05};  Bloom et al.   \cite{Bloo05b}), at a  redshift of
$z=0.257$ (Prochaska et al.  \cite{Proc05b}, \cite{Proc05c}).  Optical
(Gal-Yam et  al.  \cite{Galy05b}), near-infrared (NIR;  Cobb \& Bailyn
\cite{Cobb05}),  and radio (Berger  et al.   \cite{Berg05}) transients
were detected  offset $\sim3$~kpc  from the host,  a value  similar to
that    measured   for   the    GRB~050709   afterglow    (Hjorth   et
al. \cite{Hjor05}).

The  study of  the  spectral  energy distribution  (SED)  of GRB  host
galaxies provides  information on  the star-formation rate  (SFR), the
stellar  mass  content,  the  stellar  population  age,  the  absolute
luminosity, and the extinction.  In the present paper we construct the
SED of the GRB~050724 host  galaxy and compare its properties with the
ones reported  for the hosts of  the long-duration GRBs  (Chary et al.
\cite{Char02};   Christensen,  Hjorth,  \&   Gorosabel  \cite{Chri04};
Gorosabel et  al.  \cite{Goro05a} and  references therein). Throughout
this  paper  we   assume  a  cosmology  where  $\Omega_{\Lambda}=0.7$,
$\Omega_{M}=0.3$,  and $H_0=65$ km  s$^{-1}$ Mpc$^{-1}$.   Under these
assumptions the luminosity distance of GRB~050724 is $d_l=1\,400$ Mpc,
and the angular scale corresponds to 4.3 kpc/arcsec.

Section~\ref{Observations}  reports the  observations carried  out for
the host galaxy of GRB~050724, and Sect.~\ref{method} explains the SED
construction  method  and  Sect.~\ref{results}  reports  our  results.
Section~\ref{Discussion}  explains  our  statistical  analysis,  while
Sect.~\ref{Conclusions}  summarises the  final conclusions.   Our work
represents  the  first statistical  attempt  to  compare the  dominant
stellar ages of short and long GRB hosts.

\section{Observations and data analysis}
\label{Observations}

The data  were reduced with  IRAF\footnote{IRAF is distributed  by the
  National Optical Astronomy Observatories,  which are operated by the
  Association of  Universities for Research in  Astronomy, Inc., under
  cooperative  agreement   with  the  National   Science  Foundation.}
  following standard procedures.   Table~\ref{table1} displays the log
  of  our observations.  The $UBVR$-band  frames were  taken  with the
  Andaluc\'{\i}a Faint Object and  Spectrograph Camera (ALFOSC) at the
  2.5~m Nordic  Optical Telescope (2.5NOT).  The ALFOSC  detector is a
  2\,048$\times$2\,048  Thinned Loral  CCD providing  a field  of view
  (FoV)  of   $6\farcm5  \times  6\farcm5$   and  a  pixel   scale  of
  $0\farcs189$/pix.   The optical calibration  was performed  with the
  1.5 m telescope at the Observatorio de Sierra Nevada (OSN) observing
  a Landolt  field at an airmass  similar to that of  the GRB (Landolt
  \cite{Land92}).

The  $J_sHK_s$-band observations  were  carried out  with  the 3.58  m
Telescopio   Nazionale  Galileo   (3.6TNG)  equipped   with   NICS,  a
1\,024$\times$1\,024  HgCdTe  CCD that  provides  a  FoV of  $4\farcm2
\times 4\farcm2$ and a pixel scale of $0\farcs25$/pix. The calibration
was achieved  using eight  stars of the  2MASS catalogue that  yield a
zero  point error  of 0.03  mag in  the $J_sHK_s$  bands.   An earlier
$J$-band observation was performed on 25.868 UT July 2005 with the 3.5
m Calar  Alto telescope  (3.5CAHA) equipped with  OMEGA$_{2000}$.  The
3.5CAHA  observation   was  only  used  for   checking  the  afterglow
contamination.  Given its  large error and the existence  of much more
accurate $J_s$-band measurements  on 4.872 UT and 12.887  UT Aug 2005,
the 3.5CAHA  $J$-band magnitude was not  used to construct  the SED of
the host  galaxy.  We complemented  our $UBVRJ_sHK_s$-band photometric
data  points with  an $I$-band  magnitude  ($I=18.68\pm0.16$) obtained
from  the  average of  the  measurements  reported  by Berger  et  al.
(\cite{Berg05}) and Cobb \& Bailyn (\cite{Cobb05}).

The observations of the host presented in Table~\ref{table1} were made
at least  11.3 days after  the GRB (except  the 3.5CAHA data  taken on
25.868   UT  July  2005).    Using  $K$-band   data,  Berger   et  al.
(\cite{Berg05}) imposed an upper  limit on the afterglow's decay index
of  $\alpha  <  -1.9$.   Thus   assuming  a  maximum  decay  index  of
$\alpha=-1.9$  we  estimated  an  afterglow contribution  of  at  most
$\sim0.02\%$ to the  total light on 4.892 UT Aug  2005 (our first host
galaxy  $K_s$-band  observing   epoch).   In  addition,  the  $J$-band
magnitude measured with the 3.5CAHA telescope only 1.34 days after the
gamma-ray event is consistent  with the 3.6TNG $J_s$-band measurements
taken  on 4.872 UT  and 12.887  UT Aug  2005, supporting  (despite its
large  magnitude error)  a  low afterglow  contamination  on our  host
magnitudes as displayed in Table~\ref{table1}.

\section{Modelling the optical/NIR SED}
\label{method}

The     synthetic     SED     analysis     is     based     on     the
HyperZ\footnote{http://webast.ast.obs-mip.fr/hyperz/} code (Bolzonella
et  al.  \cite{Bolz00}).   Eight synthetic  spectral types  were used:
Starburst (Stb),  Ellipticals (E), Lenticulars (S0),  Spirals (Sa, Sb,
Sc and Sd), and Irregular  galaxies (Im).  Each spectral type has 
  an associated  star-formation rate  temporal history SFR($t$)
and a characteristic time  scale SFR $\propto \exp{(-t/\tau)}$ ($\tau$
ranges  from  0  for  Stb  to  $\infty$ for  Im  galaxies).   For  the
generation of  the synthetic  templates, three initial  mass functions
(IMFs) were considered: Salpeter (\cite{Salp55}; S55), Miller \& Scalo
(\cite{Mill79};   MS79),  and   Scalo   (\cite{Scal86};  S86).    Four
extinction   laws   were  taken   into   account:   Calzetti  et   al.
(\cite{Calz00};  suitable for  Stbs), Seaton  (\cite{Seat79};  for the
Milky Way,  MW), Fitzpatrick (\cite{Fitz86}; for  the Large Magellanic
Cloud,  LMC)  and  Pr\'evot  et  al.  (\cite{Prev84};  for  the  Small
Magellanic Cloud, SMC).
 
For  the construction  of  the synthetic  SEDs  two metallicity  ($Z$)
scenarios were  considered: i) a  static metallicity model  with fixed
metallicity values  and ii) a  dynamic close-box model  (Bolzonella et
al.  \cite{Bolz00}).  Model ii)  considers the galaxy metal enrichment
due to  ejection of  heavy elements by  each stellar  generation.  For
Model i) two fixed  metallicity values were adopted: $Z=Z_{\odot}$ and
$Z=Z_{\odot}/5$, where $Z_{\odot}=0.02$.   The metallicities of i) and
ii)  diverge  with time  defining  different  synthetic templates  for
evolved stellar populations, especially  for large $\tau$ values (i.e.
for E galaxies).

In  order to  derive the  corresponding effective  wavelengths  and AB
magnitudes,  we convolved  each  filter transmission  curve, plus  the
corresponding  CCD  efficiency  curve,  with the  Vega  spectrum  (see
Table~\ref{table1}).  For the construction of the photometric SED, all
the magnitudes were corrected  for Galactic foreground reddening along
the   GRB   line   of   sight   (Schlegel   et   al.    \cite{Schl98};
$E(B-V)=0.613$), assuming a typical MW extinction law (Cardelli et al.
\cite{Card89}). Other  Galactic $E(B-V)$ values were  also tested (see
Sect.~\ref{results}).

\begin{figure}[t]
\begin{center}
\resizebox{\hsize}{!}{\includegraphics[bb=25 45 604 595]{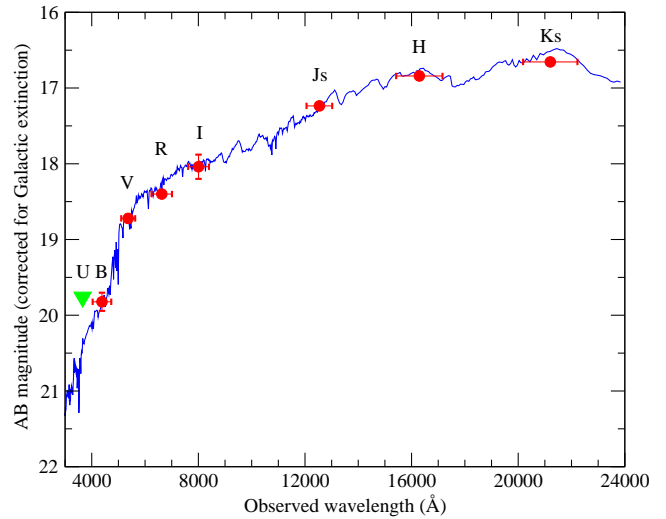}}
\caption{\label{fig1} The optical/NIR SED of the GRB~050724 host.  The
  circles represent the measured magnitudes of the $BVRIJ_sHK_s$ bands
  in   the   AB   system   and  corrected   for   Galactic   reddening
  ($E(B-V)=0.613$; Schlegel et al.  \cite{Schl98}). The triangle shows
  the  $U$-band upper  limit.   The plot  shows  the satisfactory  fit
  achieved ($\chi^2/d.o.f.  = 0.90$) with an S86 IMF, an MW extinction
  law, and  evolving metallicity.  The derived  stellar population age
  is 2.6 Gyr.}
\end{center}
\end{figure}

\section{Results}
\label{results}

Table~\ref{table1} shows the  optical/NIR magnitudes not corrected for
Galactic and  host reddening.  The  magnitudes of the host  galaxy are
based on the {\tt MAG\_AUTO}  magnitude given by SExtractor (Bertin \&
Arnouts \cite{Bert96}).  The errors in the magnitudes account for both
the zero point and the statistical errors.

Given  that the  spectroscopic redshift  is  known, we  only used  the
photometric redshift as a sanity  check for our SEDs.  The accuracy of
the photometric redshift determination  strongly depends on the number
of bands  bracketing the  rest-frame $\sim$4\,000~\AA ~break.   In our
case only the uncertain $B$-band  magnitude is at a wavelength that is
shorter than 4\,000$\times(1+z)$~\AA, providing a poor coverage of the
Balmer break position.   Nonetheless the inferred photometric redshift
($z=0.22\pm0.03$)  is  reasonably  consistent with  the  spectroscopic
redshift  ($z=0.257$).  Thus  hereafter  we fix  the  redshift of  the
templates  at  $z=0.257$.   Figure~\ref{fig1}  shows  our  photometric
points plotted  over the  SED solution obtained  for an  MW extinction
law, an S86 IMF, and evolving metallicity.

We estimate a mean $M_B=-20.7$, corresponding to $L_B\sim1.1L^{\star}$
(Schechter  \cite{Sche76}),   which  is  fainter   than  the  $K$-band
luminosity  ($L_K  \sim 1.6  L^{\star}$)  inferred  by  Berger et  al.
(\cite{Berg05}).  We  note the uncertainty  in our $M_B$ value  due to
the  high  Galactic reddening.   The  derived  stellar population  age
ranges from 2.6  to 6.5 Gyr depending on the IMF.   The S86 IMF yields
systematically   better   fits   than    the   MS79   and   S55   IMFs
($\chi^2/d.o.f\sim1$  for S86  vs.  $\chi^2/d.o.f  > 5$  for  MS79 and
S55).  The S86 IMF is  representative of stellar populations with many
solar   mass  stars  and   few  massive   stars  (Bolzonella   et  al.
\cite{Bolz00}).  The  best fit ($\chi^2/d.o.f=0.90$)  is obtained with
an S86  IMF and an age  of 2.6 Gyr (see  Fig.~\ref{fig1}).  This value
agrees  with (and  possibly improves)  the age  lower limit  of  1 Gyr
estimated  by Berger  et  al.  (\cite{Berg05})  and  Prochaska et  al.
(\cite{Proc05a}), which was based on the absence of strong H$_{\beta}$
absorption features.

The  optical spectra  reported  for GRB~050724  do  not show  emission
features   (Berger   et   al.    \cite{Berg05};   Prochaska   et   al.
\cite{Proc05a}), so a direct measurement of the host galaxy extinction
(for  instance based  on the  Balmer decrement;  extensively  used for
long-duration GRB hosts,  as well as for GRB~050709;  Prochaska et al.
\cite{Proc05a}) was not possible.   Our optical/NIR SED fitting brings
us the  opportunity to estimate  its value and  to hint at  whether it
could  play  a  crucial  role  in the  faintness  of  the  optical/NIR
emission.   Moreover  if the  SEDs  of  short  GRB host  galaxies  are
dominated  by  old stellar  populations  (as  suggested  here), it  is
plausible that  the future spectra  of such hosts will  frequently not
show  emission lines  either.   Thus the  SED-fitting technique  might
become a very valuable tool for estimating extinctions in the hosts of
short GRBs.

In  the case  of  the host  galaxy  of GRB~050724,  the inferred  mean
extinction  value is  low  ($A_{\rm v}=0.2\pm0.2$)  regardless of  the
assumed extinction law.   This value is similar to  the low extinction
derived for the host  of GRB~050709 (Prochaska et al.  \cite{Proc05a};
Hjorth et  al.  \cite{Hjor05}). So  in principle the faintness  of the
optical/NIR afterglow cannot be  explained by the global extinction of
the host galaxy.  But we note that  in a few cases dark GRBs have been
associated   with    low   extinction   hosts    (Gorosabel   et   al.
\cite{Goro03}).   Given the high  foreground Galactic  extinction (and
its large associated uncertainty),  we ranged $E(B-V)$ from $0.613$ to
$0.265$  (as reported  by Burstein  \& Heiles  \cite{Burs82}), thereby
extending previous analyses (i.e.   Berger et al.  \cite{Berg05}) that
kept   it  fixed   at  the   value  derived   from  Schlegel   et  al.
(\cite{Schl98}).    For  $E(B-V)=0.265$   the  inferred   host  galaxy
extinction increased to $A_{\rm v} \sim 0.4$, and the dominant stellar
population age  increased to  5 Gyr.  Still  the best fit  is obtained
with an elliptical galaxy, an S86 IMF, and evolving metallicity.

The rest-frame  ultraviolet (UV)  flux at 2\,800~\AA  ~(3\,520~\AA~ in
the observer  frame) can provide  an estimate of  the SFR in  a galaxy
(Kennicutt  \cite{Kenn98}).   It  has  been extensively  used  in  SFR
calculations  of  long GRB  hosts  (Christensen,  Hjorth \&  Gorosabel
\cite{Chri04}), but for the host galaxy  at hand it might result in an
overestimate.  This is due to the fact that massive galaxies dominated
by old stellar populations (age $>$ 1 Gyr) often show a UV excess (the
UV  upturn,  see  Yi  \&  Yoon  \cite{Yi04}  and  references  therein)
originating  in the  helium-burning  stars in  the horizontal  branch.
Hence the proportionality between the  UV luminosity and the SFR might
not  be  valid  for  the   host  of  GRB~050724.  When  the  Kennicutt
(\cite{Kenn98}) relation is applied to  our UV flux we obtain an
SFR value of $0.5 - 5 M_{\odot}$ yr$^{-1}$ (corrected for Galactic and
host galaxy extinction), well above  the SFR upper limits reported for
GRB~050724  (SFR   $<  0.02   M_{\odot}$  yr$^{-1}$,  Berger   et  al.
\cite{Berg05};  SFR $<  0.05  M_{\odot}$ yr$^{-1}$,  Prochaska et  al.
\cite{Proc05a};  both  based on  optical  spectra).   This fact  might
indicate the presence of an  old stellar population. But the uncertain
Galactic reddening  and inaccurate SED extrapolation to  the UV domain
(which is  highly template dependent) do  not permit us to  claim a UV
upturn for our host.

\section{Discussion: statistical analysis}
\label{Discussion}

\begin{figure}[t]
\begin{center}
\resizebox{\hsize}{!}{\includegraphics[bb= 35 35 707 590]{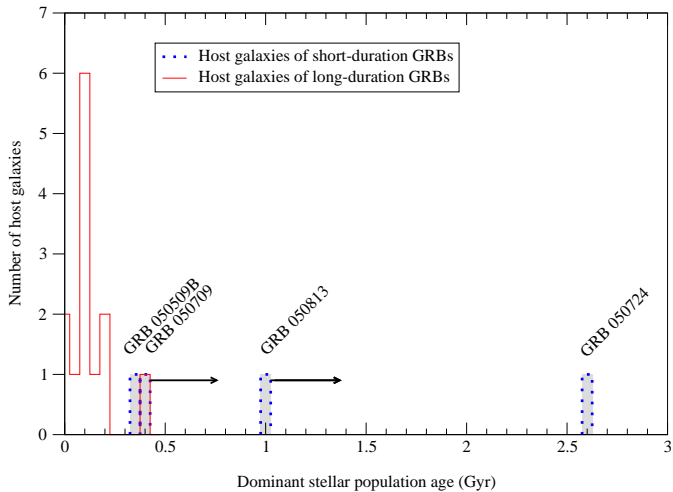}}
\caption{\label{fig2}  The   distribution  of  the   dominant  stellar
  population ages  for short- and long-duration GRB  hosts.  As shown,
  host galaxies of short GRBs (filled bins) tend to have older stellar
  populations than  long GRB hosts.   According to the WMWW  test, the
  probability that the two samples  (short and long GRB host galaxies)
  share the  same parent median  age is only  $2.5\times10^{-3}$.  The
  arrows  indicate  that the  stellar  ages  assumed for  GRB~050509B,
  GRB~050709, and GRB~050813 are probably underestimated.}
\end{center}
\end{figure}

Figure~\ref{fig2}  shows  the  distribution  of the  dominant  stellar
population  ages for  two samples  of long  and short  GRB  hosts. The
sample of long-duration GRB  host galaxies contains $N_l=13$ galaxies,
based on Christensen,  Hjorth, \& Gorosabel (\cite{Chri04}), Gorosabel
et  al.   (\cite{Goro05a}), and  Chary  et  al.  (\cite{Char02}).   We
considered a  short-duration GRB sample containing only  $N_s= $4 host
galaxies  (GRB~050509B, GRB~050709,  GRB~050813, and  GRB~050724).  To
consider  this  sample we  compiled  the  stellar  ages for  the  host
galaxies of  GRB~050509B and GRB~050709 and roughly  estimated the age
of the  host of GRB~050813,  which has not  been reported to  date. We
used the Kolmogorov-Smirnov (K-S) test  to compare the stellar ages of
the long and short GRB host  samples.  The sample of $N_s=4$ short GRB
hosts,  although it  falls short,  is close  to the  K-S applicability
limit ($\frac{N_s N_l}{N_s+N_l}=3.06$; Press et al. \cite{Pres92}).
  
Castro-Tirado et al.  (\cite{Cast05})  reported a rather young stellar
population age (0.36~Gyr)  for the host galaxy of  GRB~050509B using a
photometric $BRIJHK$-band SED. This age  estimate is based on the size
of the 4\,000~\AA~ jump, which requires it to be well bracketed by the
SED photometric  bands.  For the redshift  of GRB~050509B ($z=0.225$),
the SED coverage  was possibly too reduced to  determine the full size
of the 4\,000~\AA~  spectral step.  In fact the  4\,000~\AA~ break (at
$\sim$4\,900~\AA~  for  $z=0.225$)  falls  on  the  outskirts  of  the
$B$-band,  the only filter  placed on  the blue  side of  the spectral
break (see  Fig.~2 of Castro-Tirado et  al. \cite{Cast05}).  Therefore
it is plausible that the size  of the 4\,000~\AA~ break, and hence the
GRB~050509B host  stellar age, were underestimated.   This agrees with
the lack  of star-formation activity  and emission lines found  in the
GRB~050509B spectrum (Bloom et al.  \cite{Bloo05a}) and with the 1~Gyr
age upper  limit reported by Prochaska et  al.  (\cite{Proc05a}).  But
for the statistical comparison we  decided to use the youngest stellar
age of 0.36~Gyr reported for the host galaxy of GRB~050509B.

A  similar age  underestimate  cannot  be discarded  for  the host  of
GRB~050709, for which Hjorth et al.  (\cite{Hjor05}) estimate a rough,
dominant  stellar age  of  0.4~Gyr  based on  a  $BVRI$-band SED.   In
contrast,  Covino  et  al.    (\cite{Covi05b})  report  an  older  age
($\gtrsim$1~Gyr)  based on  spectral  data.  For  the  host galaxy  of
GRB~050709, as for the GRB~050509B host, we assumed an age lower limit
of 0.4~Gyr.

The X-ray error box of GRB~050813  is centred in a cluster of galaxies
(similar  to   the  host  galaxy  of  GRB~050509B,   Pedersen  et  al.
\cite{Pede05})  at  $z=0.722$, and  contains  two  red ($K\sim19,  R-K
\gtrsim  4$;   Gladders  et  al.   \cite{Glad05};   Prochaska  et  al.
\cite{Proc05a})  elliptical  galaxies  consistent  with  the  redshift
cluster (B  and C, following  Gorosabel et al.  \cite{Goro05b})  and a
fainter one with unknown redshift (B$^{\star}$, following Prochaska et
al.   \cite{Proc05a}).   In  the   OSN  $I$-band  images  reported  in
Gorosabel  et al.   (\cite{Goro05b})  the  galaxies B  and  C show  an
angular  size of $\sim3-4^{\prime\prime}$,  corresponding to  a linear
diameter of  $\sim25-35$~kpc at $z=0.722$.   The faintness of  the two
galaxies in the OSN $I$-band images and the proximity of a bright star
($R\sim 15.2$,  $\sim10^{\prime\prime}$ to the  southeast) prevented a
study of  their surface brightness  profile.  Given the  linear sizes,
colours, and locations  of the galaxies B and C (in  the centre of the
galaxy  cluster) both  very  likely correspond  to massive  elliptical
galaxies,   as   suggested  by   other   authors   (Gladders  et   al.
\cite{Glad05};  Prochaska  et al.   \cite{Proc05a}).   Cimatti et  al.
(\cite{Cima02}) studied  a sample of  78 galaxies with  magnitudes ($K
\lesssim 19.2$), colours ($R-K \gtrsim  5$), and redshifts ($0.7 < z <
1.5$) that are  similar to the two galaxies  located in the GRB~050813
error box,  deriving a minimum stellar  age of $\sim3$~Gyr  for the 78
galaxies.  Thus we  assumed a conservative age lower  limit of $1$~Gyr
for the GRB~050813 host stellar population.

The  K-S test gives  a probability  of $P=3.5\times10^{-3}$  that both
samples  share the  same  parent age  distribution.  This  statistical
comparison strongly  suggests that  short-duration GRBs occur  in host
galaxies with  older stellar populations than the  progenitors of long
GRBs.   We note  that the  K-S applicability  criterion  (described by
$\frac{N_s N_l}{N_s+N_l} \geq 4$)  is not formally satisfied, since it
is currently  limited by the  low $N_s$ value  (but not by  $N_l$). In
order to strictly meet the  K-S applicability conditions, it remains a
priority to increase the short GRB host sample to $N_s \ge 6$.

To  check the  impact  of the  reduced  samples on  the  K-S test,  we
complemented  the analysis  with the  W  Mann-Withney-Wilconxon (WMWW)
test   (Wall   \&   Jenkins   \cite{Wall03};   Siegel   \&   Castellan
\cite{Sieg88}).  The WMWW  test is more appropriate than  the K-S test
for  looking at  the  differences  in the  central  tendencies of  two
samples, which seems  to be the case for our  two host galaxy samples;
see in  Fig.~\ref{fig2} how short GRB  hosts tend to  have older ages.
This test  is non-parametric (unlike other tests,  i.e.  the T-student
test, it does not make any assumptions about the parent distribution),
and  is  suitable for  comparing  samples  with  a reduced  number  of
objects.  According to the WMWW test, the probability that the two age
samples  share   a  parent  median  is   only  $P=2.5\times  10^{-3}$,
supporting the K-S results.

For  long  GRB host  galaxies  there  is  no correlation  between  the
redshift and the age of  the stellar population.  Thus the conclusions
are  unaltered  if the  sample  of short  bursts  is  compared to  any
subsample of long GRB hosts constructed based on a redshift criterion.

\section{Conclusions}
\label{Conclusions}

The optical/NIR SED  constructed for the host galaxy  of GRB~050724 is
reproduced   well  ($\chi^2/d.o.f.=0.90$)  by   a  massive   ($M  \sim
5\times10^{10} M_{\odot}$)  elliptical galaxy with  an evolved stellar
population (age $\sim 2.6$ Gyr).  The best fit is obtained with an S86
IMF, which suggests a stellar population dominated by solar-mass stars
and with few  massive stars.  If the metallicity of  the host is fixed
to a  value constant  in time, no  satisfactory SED fits  are achieved
($\chi^2/d.o.f.  > 19$ for  $Z=Z_{\odot}$).  Satisfactory SED fits are
obtained  only if  the host  galaxy metallicity  increases  with time.
These  findings  suggest that  GRB~050724  is  the  consequence of  an
evolved stellar population consistent  with being caused by the merger
of two compact objects, a historical prediction of several theoretical
models.   It is  interesting to  note that  this conclusion  joins the
ranks of  those recently accumulated for  other, well-localised, short
GRBs.  In a  broader context our comparative study,  based on 4 short-
(including  GRB~050724) and 13  long-duration GRBs,  strongly suggests
that the  dominant stellar populations of  short GRB hosts  tend to be
older than the ones in host galaxies of long GRBs.

A scenario as  simplified as the one presented here is  likely to be a
naive  description of a  complex reality.   Much as  not all  long GRB
hosts are associated with  young, subluminous, starburst galaxies (the
spiral hosts of GRB~980425 and GRB~990705 are clear examples; Fynbo et
al.  \cite{Fynb00},  Le~Floc'h et  al.  \cite{LeFl02}), not  all short
GRBs  are  linked  to  old,  elliptical  galaxies  (the  case  of  the
GRB~050709 host galaxy). The limited sample of short GRB hosts implies
that  the  above  conclusions  are  still  governed  by  small  number
statistics,  and  a  detailed   picture  cannot  be  drawn  yet.   The
identification  of   new,  short-duration,  GRB   host  galaxies  will
definitively allow for more discriminating statistical comparisons.

\section*{Acknowledgments}

This  research  was  partially  supported  by  Spain's  Ministerio  de
Educaci\'on  y  Ciencia  through programmes  ESP2002-04124-C03-01  and
AYA2004-01515 (including  FEDER funds).  JMCC  gratefully acknowledges
partial support from the Instrumentcenter for Dansk Astrofysik and the
NBI's   International   Ph.D.   School   of   Excellence.   We   thank
M.~Cervi\~{n}o  and  J.F.   G\'omez  for  helpful  discussion  of  the
statistical analysis.  We  are grateful to C.  Coutures  and J. Hjorth
for their invaluable help. We also wish to thank our anonymous referee
for useful comments.

\end{document}